\documentclass[amsmath,amssymb,aps,pra,superscriptaddress,twocolumn,floatfix]{revtex4}

\usepackage{graphicx}
\makeatletter
\def\justified{
	\let\\\@normalcr
	\@rightskip\z@skip \rightskip\@rightskip
	\leftskip\z@skip
	\parindent 0em\relax
	\setlength{\parfillskip}{0pt plus 1fil}}
\usepackage{amsmath}
\usepackage{braket}
\usepackage{units}
\usepackage{ragged2e}
\usepackage[colorlinks,urlcolor=blue ,citecolor=blue ,linkcolor=blue ]{hyperref}
\usepackage{xcolor}
\usepackage{nicefrac} 
\usepackage{lipsum}
\usepackage{nameref}
\usepackage{hyperref}
\usepackage{textcomp} 
\usepackage{urwchancal}
\usepackage{upgreek} 
\usepackage{amsmath} 
\usepackage{here}
\usepackage{siunitx}
\usepackage{booktabs}
\usepackage{longtable}

\newcommand{\gJ}{g_J}
\newcommand{\gJp}{g_{J'}}
\newcommand{\gs}{$\ket{g}$}
\newcommand{\es}{$\ket{e}$}

\begin{document}

\title{Observation of a narrow inner-shell orbital transition in atomic erbium at 1299\,nm}
\author{A.\,Patscheider}
\thanks{These authors contributed equally to this work.}
\affiliation{Institut f\"ur Experimentalphysik, Universit\"at Innsbruck, Technikerstra{\ss}e 25, 6020 Innsbruck, Austria}

\author{B.\,Yang}
\thanks{These authors contributed equally to this work.}
\affiliation{Institut f\"ur Experimentalphysik, Universit\"at Innsbruck, Technikerstra{\ss}e 25, 6020 Innsbruck, Austria}

\author{G.\,Natale}
\affiliation{Institut f\"ur Experimentalphysik, Universit\"at Innsbruck, Technikerstra{\ss}e 25, 6020 Innsbruck, Austria}

\author{D.\,Petter}
\altaffiliation[Present address:]{ Optical Materials Engineering Laboratory, Department of Mechanical and Process Engineering, ETH Zurich, 8092 Zurich, Switzerland.}
\affiliation{Institut f\"ur Experimentalphysik, Universit\"at Innsbruck, Technikerstra{\ss}e 25, 6020 Innsbruck, Austria}

\author{L.\,Chomaz}
\altaffiliation[Present address:]{ Physikalisches Institut, University of Heidelberg, 69120 Heidelberg, Germany.}
\affiliation{Institut f\"ur Experimentalphysik, Universit\"at Innsbruck, Technikerstra{\ss}e 25, 6020 Innsbruck, Austria}

\author{M.\,J.\,Mark}
\affiliation{Institut f\"ur Experimentalphysik, Universit\"at Innsbruck, Technikerstra{\ss}e 25, 6020 Innsbruck, Austria}
\affiliation{Institut f\"ur Quantenoptik und Quanteninformation, \"Osterreichische Akademie der Wissenschaften, Technikerstra{\ss}e 21a, 6020 Innsbruck, Austria}

\author{G.\,Hovhannesyan}
\affiliation{Laboratoire Interdisciplinaire Carnot de Bourgogne, CNRS, Universit\'e de Bourgogne Franche-Comt\'e, 21078 Dijon, France}

\author{M.\,Lepers}
\affiliation{Laboratoire Interdisciplinaire Carnot de Bourgogne, CNRS, Universit\'e de Bourgogne Franche-Comt\'e, 21078 Dijon, France}

\author{F.\,Ferlaino}
\affiliation{Institut f\"ur Experimentalphysik, Universit\"at Innsbruck, Technikerstra{\ss}e 25, 6020 Innsbruck, Austria}
\affiliation{Institut f\"ur Quantenoptik und Quanteninformation, \"Osterreichische Akademie der Wissenschaften, Technikerstra{\ss}e 21a, 6020 Innsbruck, Austria}

\begin{abstract}
    We report on the observation and coherent excitation of atoms on the narrow inner-shell orbital transition, connecting the erbium ground state $[\mathrm{Xe}] 4f^{12} (^3\text{H}_6)6s^{2}$ to the excited state $[\mathrm{Xe}] 4f^{11}(^4\text{I}_{\nicefrac{15}{2}})^05d (^5\text{D}_{\nicefrac{3}{2}}) 6s^{2} (\nicefrac{15}{2},\nicefrac{3}{2})^0_7$. This transition corresponds to a wavelength of \SI{1299}{nm} and is optically closed. We perform high-resolution spectroscopy to extract the $\gJ$-factor of the $1299$-nm state and to determine the frequency shift for four bosonic isotopes. We further demonstrate coherent control of the atomic state and extract a lifetime of \SI{178(19)}{ms} which corresponds to a linewidth of \SI{0.9(1)}{Hz}. The experimental findings are in good agreement with our semi-empirical model. In addition, we present theoretical calculations of the atomic polarizability, revealing several different magic-wavelength conditions. Finally, we make use of the vectorial polarizability and confirm a possible magic wavelength at \SI{532}{nm}.
\end{abstract}

\date{\today}

\maketitle

\section{Introduction}

Ultra-narrow atomic transitions are an extremely powerful resource for high-precision measurements and for controlling and manipulating atoms on a quantum level~\cite{Zhang2016}. Prominent examples are clock transitions in alkaline-earth-like atoms~\cite{Ludlow2015,McGrew2018,Schaefer2020}. The small spectral linewidth of these transitions enables the high-resolution detection of energy shifts on very fine scales. This unique property made it possible, e.\,g.\,to observe SU(N)-symmetric interactions in both, ytterbium and strontium~\cite{Scazza2014,Zhang2014}. An additional important avenue paved by narrow transitions is the optical manipulation and coherent control of ultracold atoms. The tuning of the inter-particle interactions using optical Feshbach resonances has been demonstrated and benefits from the narrow linewidth due to the suppressed photon scattering rate~\cite{Fatemi2000,Theis2004,Blatt2011,Saha2014,Nicholson2015}. Coherent control enabled the creation of ultracold molecules via Raman state transfer~\cite{Winkler2007,Danzl2008,Ni2008,Reinaudi2012}, the preparation of the atoms in different nuclear spin configurations~\cite{Scazza2014,Zhang2014}, and the creation of spin-orbit coupled quantum gases~\cite{Hasan2010,Galitski2013,Livi2016,Kolkowitz2017}. Finally, the coherent excitation allows for the realization of quantum computation and quantum simulations, e.\,g.\, with neutral atoms loaded into optical lattices~\cite{Daley2008,Gorshkov2009,Heinz2020}.

Atomic species of the lanthanide family are multi-valence electron atoms and possess a special electron configuration, a so-called submerged shell, in which the $6s$ sub-shell is filled, while the lower-lying $4f$ or $5d$ sub-shells are open, being partially unoccupied. This leads to a large variety of optical transitions in these elements, whose linewidths range from tens of \si{\micro Hz} to tens of \si{MHz}~\cite{Ban2005,Dzuba2010,Kozlov2013}. In contrast to alkaline-earth-like atoms, which do not carry a magnetic moment in their ground state, a selection of lanthanides allow for the combination of a narrow transition with a large magnetic moment. While narrow and ultra-narrow transitions have been extensively studied in alkaline-earth and ytterbium atoms, only little is known for the other elements of the lanthanide series. Some spectroscopic studies have been carried out for dysprosium~\cite{Petersen2020} and thulium~\cite{Golovizin2019,Tregubov2020}.

\begin{figure}
    \centering
    \includegraphics[width = 0.5\textwidth]{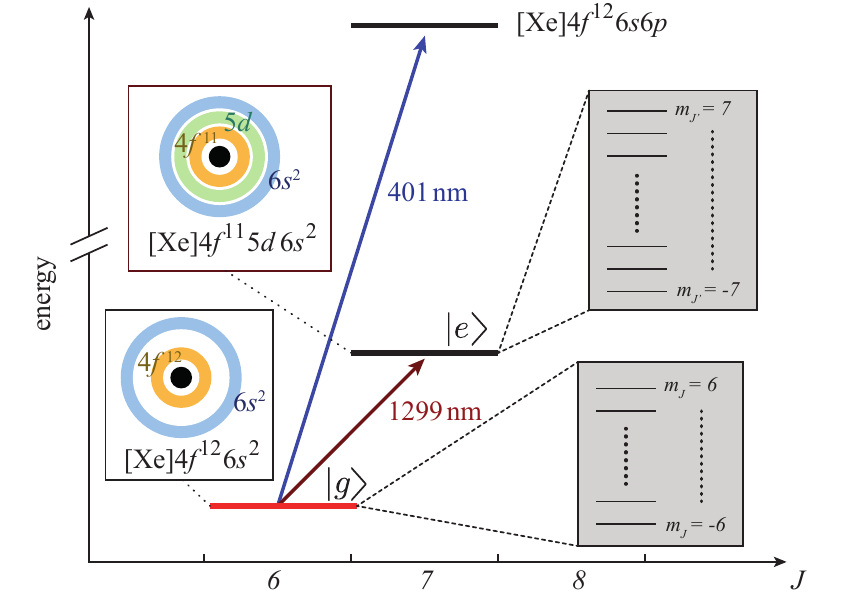}
    \caption{Schematic level scheme illustrating the $[\mathrm{Xe}] 4f^{12} (^3\text{H}_6)6s^{2}$  $\to$ $[\mathrm{Xe}] 4f^{11}(^4\text{I}_{\nicefrac{15}{2}})^05d (^5\text{D}_{\nicefrac{3}{2}}) 6s^{2} (\nicefrac{15}{2},\nicefrac{3}{2})^0_7$ inner-shell orbital transition at \SI{1299}{nm} ($\ket{g}$ $\to \ket{e}$) and the $[\mathrm{Xe}] 4f^{12} (^3\text{H}_6)6s^{2}$ $\to$ [Xe]$4f^{12}(^3\text{H}_6)6s6p (^1\text{P}_1) (6,1)^0_7$ transition at \SI{401}{nm}, used for absorption imaging. The horizontal lines indicate the energy levels for \gs~(blue, even parity), \es~(red, odd parity), and the state at \SI{401}{nm} (black, odd parity). [Xe] stands for the electron configuration of xenon. The insets illustrate the electron configurations. The grey shaded boxes represent the Zeeman manifold for \gs~and \es. Energies are not to scale.}
    \label{fig:Fig1}
\end{figure}

For the specific case of erbium, there is a prediction of a narrow inner-shell orbital transition, which has a change in the total angular moment of $\Delta J = +1$ ($\ket{J = 6} \to \ket{J' = 7}$) and a change in the total spin of $\Delta S=1$~\cite{Ban2005}. The transition involves the excitation of a $4f$ ground-state electron to a $5d$-state; see Fig.\,\ref{fig:Fig1}. Theoretical calculations predict a linewidth of about \SI{2}{Hz}~\cite{Ban2005}, which fills a gap between ultra-narrow transitions in the \si{mHz} regime and transitions having linewidths on the order of \si{kHz}, available in alkaline-earth atoms and previously explored in lanthanide atoms. Moreover, in contrast to most narrow transitions in other atomic species, the wavelength of \SI{1299}{nm} lies within the telecom-wavelength window, which, e.\,g.\,, is advantageous for the application in quantum communication systems~\cite{Kimble2008,Reiserer2015,Wehner2018,Menon2020}. Here, we report on the experimental observation of this transition. We perform a careful experimental survey and characterization of the $1299$-nm transition, realizing the first crucial step towards extended applications, e.\,g.\, to explore novel few and many-body phenomena in dipolar or large spin systems.

We experimentally observe the transition at \SI{1299}{nm} for the bosonic isotopes $^{164}$Er, $^{166}$Er, $^{168}$Er, and $^{170}$Er and for the fermionic isotope $^{167}$Er. We perform high-resolution spectroscopy to determine the $\gJp$-factor of the excited atomic state (\es)~and to measure the frequency shift for the four bosonic isotopes. We further demonstrate coherent control of the atomic state and measure an excited-state lifetime of \SI{178(19)}{ms}. We carry out trap frequency measurements to determine the atomic polarizability of the excited state relative to the ground state with the trapping light at \SI{532.2}{nm}. As we vary the polarization of the light we take advantage of the vectorial term of the atomic polarizability and we are able to get close to a magic-wavelength condition, where the ground state (\gs) and \es~feature the same polarizability. We finally present theoretical calculations of the atomic polarizability based on a sum-over-states formula and report on several alternative options for magic wavelengths.

\section{Experimental setup}

We search for the narrow inner-shell transition by performing spectroscopic measurements on a trapped quantum-degenerate erbium gas. Our experimental procedure to create an erbium Bose-Einstein condensate (BEC) follows Ref.\,\cite{Aikawa2012}. In brief, after laser cooling in a magneto-optical trap, we load the atoms into a crossed optical dipole trap (ODT) operating at \SI{1064}{nm} and perform evaporative cooling down to quantum degeneracy. The BEC typically contains $N = \num{1}$ - $\num{3e4}$ atoms with BEC fractions ranging from \num{30} - \SI{80}{\%}, depending on the isotope choice. For the fermionic $^{167}$Er isotope, we obtain a degenerate quantum gas of $N = \num{2e4}$ atoms at a temperature of $T \approx 0.5T_F$, where $T_F$ is the Fermi temperature. During the evaporation, a homogeneous magnetic field $B$ is applied to ensure that the atomic cloud remains spin-polarized in the lowest Zeeman level $m_J = -6$ ($m_F = -\nicefrac{19}{2}$) for the bosonic (fermionic) isotopes.

The light for driving the narrow inner-shell transition is generated from an external-cavity diode laser (ECDL) operating at \SI{1299}{nm}. We determine the absolute frequency of the laser by measuring the frequency-doubled light with a calibrated wavemeter~\cite{WavemeterCalib} which has an accuracy of \SI{60}{MHz}. For our coarse spectroscopy, we use the wide tunability of the ECDL via the control of a piezoelectric element, which allows us to change the laser frequency.
Furthermore, we can narrow the laser linewidth and stabilize the frequency using a high-finesse reference cavity made of ultra-low expansion glass. The reference cavity has a free spectral range (FSR) of \SI{1.4972462(3)}{GHz} and finesse of about \num{175000}. The stabilized laser system has an Allan deviation of \num{3.1e-15} over an observation time of \SI{1}{s}. The coherence time is extracted from the phase noise power spectral density and corresponds to \SI{96}{ms}~\cite{SLS}.  We measure a linear frequency drift of the high-finesse cavity of \SI{4.34(7)}{kHz \per day}. In the experiment, we use the frequency stabilized configuration for the high-resolution spectroscopy.

\section{Coarse spectroscopy}

In the 1960s, the atomic spectra of lanthanides began to attract interest. Absorption lines were observed using King's furnace or oxyacetylene flames in the range between 650 to \SI{250}{nm}~\cite{Mossotti1964,Marquet65}. In early spectroscopic works, the configuration of the low-lying energy levels, and particularly that of the odd parity states, was not known.
The first work to identify the odd-parity level $4f^{11}5d6s^2$ – i.\,e.\,ortho-erbium ground state – as the lowest-lying configuration above \gs~was Ref.\,\cite{Spector66}. In this configuration, the angular momentum $J_1 = \nicefrac{7}{2}$ of the $4f^{11}$ core couples via $J_1 - j$ coupling to the $j$ of the $5d$ electron, which leads to a total of 10 fine structure levels with $J$ ranging from 5 to 10. The assignment of the fine structure levels has been deduced by analyzing energy differences between absorption lines using a spectroscopic-level searching algorithm. The NIST database reports the energies of the corresponding fine-structure levels, referring to unpublished measurements from mid 70’s~\cite{vanKleef1975}.
For our level of interest $\ket{e}=4f^{11}(^4\text{I}_{\nicefrac{15}{2}})^05d (^5\text{D}_{\nicefrac{3}{2}}) 6s^{2} (\nicefrac{15}{2},\nicefrac{3}{2})^0_7$, the wavenumber given by NIST~\cite{NIST2020}, not accounting for the isotope shift, is
\begin{equation}
\label{Eq:nu_NIST}
    \Bar{\nu}_\text{NIST} = \SI{7696.956}{cm^{-1}}.
\end{equation}
To the best of our knowledge, prior to this work, there has been no direct measurement of the 1299-nm transition. Our ultracold quantum gas provides a new opportunity to observe and characterize this transition.

We start our search of the line by performing a coarse spectroscopy over a broad frequency range around $\Bar{\nu}_\text{NIST}$ (corresponding to $\nu_\text{NIST} = \SI{230.738}{THz}$).
After preparing an optically-trapped ultracold erbium gas in \gs, we shine the 1299-nm spectroscopy light on the sample. The spectroscopy beam has a peak intensity of $I_\mathrm{peak}\approx\SI{0.8}{W \per cm^2}$ and a $\nicefrac{1}{e^2}$-waist of about $\SI{110}{\micro m}$. The irradiation time is \SI{100}{ms}, during which we sweep the laser frequency with an amplitude of about $\pm \SI{40}{MHz}$. After irradiation, we release the atoms from the trap for a free expansion of \SI{30}{ms}.
We record the number of remaining \gs~atoms by performing standard absorption imaging using resonant light at \SI{401}{nm} (see Fig.\,\ref{fig:Fig1}).

\begin{figure}
    \centering
    \includegraphics[width = 0.5\textwidth]{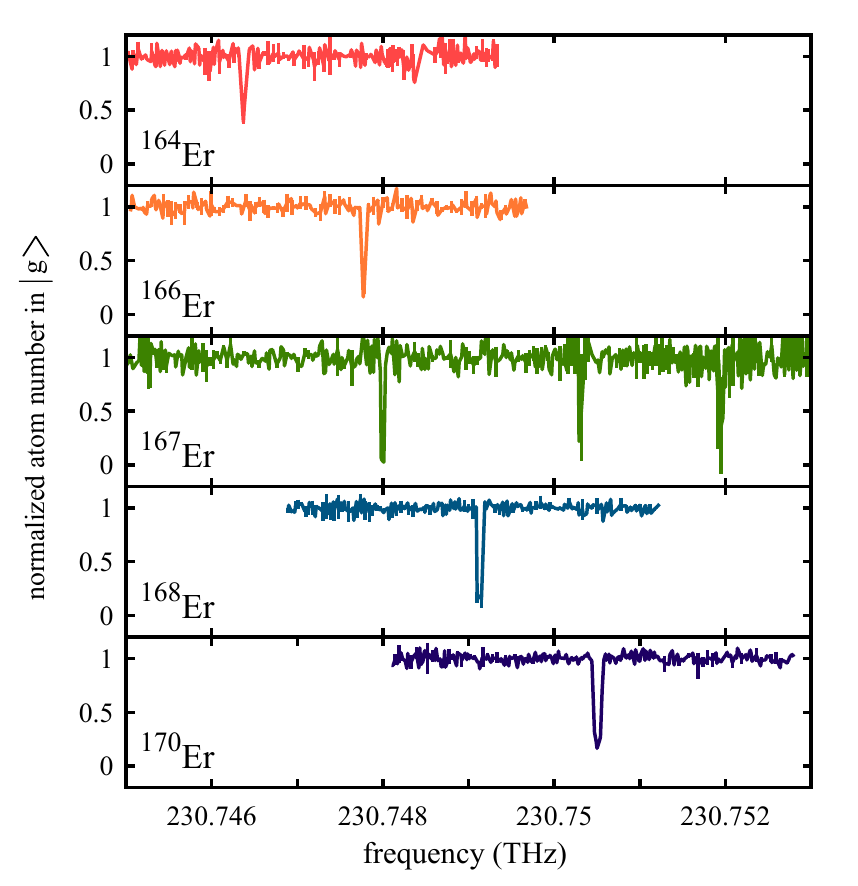}
    \caption{Coarse spectroscopy results for the four bosonic isotopes $^{164}$Er, $^{166}$Er, $^{168}$Er, and $^{170}$Er at $B = \SI{1.355(5)}{G}$ and for the fermionic isotope $^{167}$Er at $B = \SI{0.52(5)}{G}$. The normalized atom number in \gs~is plotted versus the laser frequency, which is controlled by the piezoelectric element of the ECDL. The atom number is normalized to the moving median formed by 21 data points.}
    \label{fig:FigPiezo}
\end{figure}

We record the absorption spectrum by repeating the measurement over a wide frequency range with a step size of \SI{40}{MHz}.
Figure\,\ref{fig:FigPiezo} summarizes our results for the four most abundant bosonic erbium isotopes ($^{164}$Er, $^{166}$Er, $^{168}$Er, and $^{170}$Er) and for the fermionic $^{167}$Er isotope.
As expected from their zero nuclear spin ($I=0$), each bosonic isotope exhibit just one resonant absorption line, detected as a sharp dip in the number of \gs~atoms when varying the $1299$-nm laser frequency.
For the fermionic  $^{167}$Er isotope, possessing a hyperfine structure ($I = 7/2$), we identify three resonances. We attribute the three resonances to the transitions $\ket{F = 19/2} \to \ket{F' = 21/2}$, $\ket{F = 19/2} \to \ket{F' = 19/2}$, and $\ket{F = 19/2} \to \ket{F' = 17/2}$, respectively. Notably, at the resonance positions, the population of \gs~reaches values below \num{0.5}, indicating an underlying loss mechanism, such as heating from a reduced trapping potential for \es~or an interaction-based loss processes.

\section{High-resolution spectroscopy}

Thanks to the coarse spectroscopy measurements, we can now restrict the frequency region of interest and perform a spectroscopy survey with a much higher spectral resolution and lower laser intensity, allowing also to resolve the magnetic Zeeman sub-levels.

\begin{figure}
    \centering
    \includegraphics[width = 0.5\textwidth]{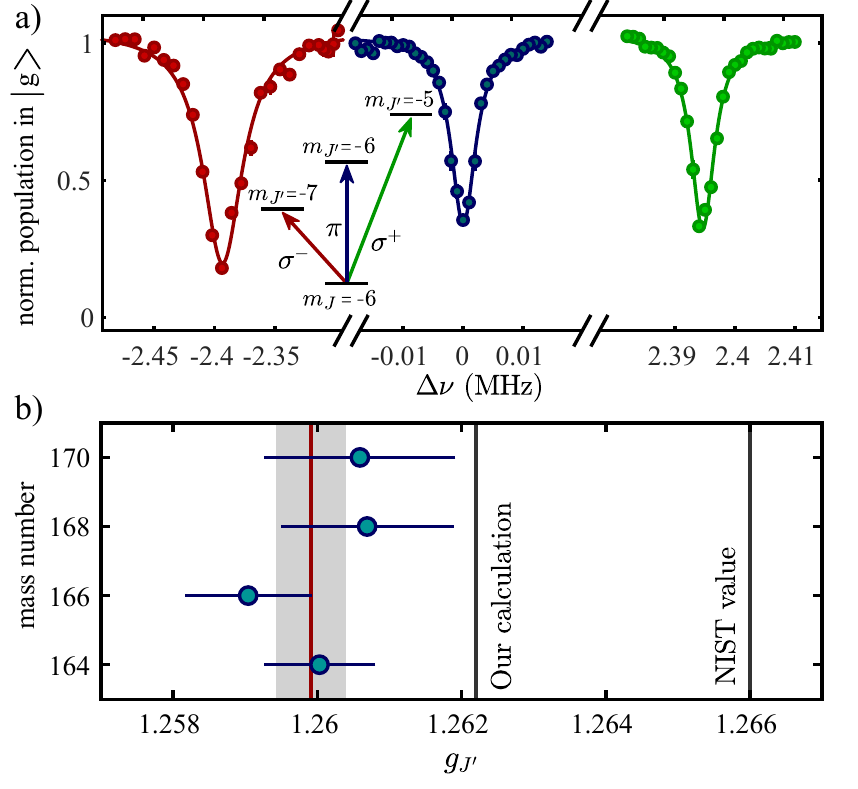}
    \caption{(a) High-resolution spectroscopy of $^{168}$Er at $B =\SI{1.358(2)}{G}$, unveiling the $\sigma^-$-, $\pi$-, and $\sigma^+$-transition ($\ket{m_J = -6} \to \ket{m_{J'} = -7}$, $\ket{m_J = -6} \to \ket{m_{J'} = -6}$, and $\ket{m_J = -6} \to \ket{m_{J'} = -5}$). The normalized population in \gs~is measured against the laser frequency relative to the frequency position of the $\pi$-transition. (b) Measured $\gJp$-factor for the four bosonic isotopes for the different atomic mass numbers. The error bars denote the $1\sigma$-standard deviation. The red solid line represents the weighted mean of the experimental data and the grey shaded area corresponds to the combined standard deviation. The black solid lines represent the $\gJp$-factor calculated using our semi-empirical method and the value given in the NIST database~\cite{NIST2020}.}
    \label{fig:FigMf_andgJ}
\end{figure}

For this, we stabilize the laser frequency to the high-finesse reference cavity and then precisely tune the laser frequency using an acousto-optical modulator. The recorded absorption spectra have a step size ranging from $1$ to \SI{8}{kHz}, depending on the measurement.
Moreover, the spectroscopy is now performed with a free-falling gas to eliminate possible AC-Stark shifts, eventually caused by the ODT light at \SI{1064}{nm}.
Therefore, after sample preparation, we switch off all trapping lights, and then irradiate the sample with a \SI{1299}{nm} spectroscopy pulse of \SI{1}{ms}, corresponding to a Fourier limited linewidth of \SI{800}{Hz}.
The pulse has a peak intensity of $I_\mathrm{peak}\approx\SI{25}{mW \per cm^2}$.
To minimize the possible frequency shifts caused by the Doppler effect in a free-falling sample, the 1299-nm laser beam propagates in a plane orthogonal to the vertical direction, defined by gravity. The quantization axis, defined by our bias magnetic field, is oriented along the vertical direction. The light contains contributions from all light polarizations, such that the 1299-nm beam can induce $\sigma^+$-, $\sigma^-$-, and $\pi$-transitions; see inset in Fig.\,\ref{fig:FigMf_andgJ}(a).

Figure\,\ref{fig:FigMf_andgJ}(a) shows the ground-state population for the $^{168}$Er isotope as a function of the laser detuning, plotted with respect to the central frequency of the $\pi$-transition. We clearly observe three resonant dips in the ground-state population, corresponding to the transitions from the ground-state level $m_{J} = -6$ to the excited Zeeman sub-levels $m_{J'} = -7,\ -6\ \text{and}\ -5$. We extract the center frequency and the transition linewidth by fitting the spectroscopy signals with a Lorentzian function. The extracted linewidths are \SI{2.4(1)}{kHz} and \SI{2.5(1)}{kHz} for the $\pi$- and $\sigma^+$- transition and \SI{20(1)}{kHz} for the $\sigma^-$- transition. The different linewidths of the spectroscopy resonances can be explained by a power broadening effect, due to the different Clebsch-Gordan coefficients of the magnetic sub-levels, and the composition of the light polarization.

We use the wavemeter to determine the absolute wavenumber as
\begin{equation}
   \Bar{\nu}_{168} = \SI{7696.955(2)}{cm^{-1}}.
\end{equation}
Our measurement is consistent with the value reported in the NIST database~\cite{NIST2020} (see Eq.\,\ref{Eq:nu_NIST}). The accuracy of the absolute wavenumber is limited by the wavemeter. However, our spectroscopy measurement has a precision of about \SI{2}{kHz}, which provides the opportunity to improve the accuracy of the absolute frequency by several orders of magnitude using advanced measurement techniques, such as frequency combs~\cite{Fortier2019}.

\section{Land\'e factor for bosonic isotopes}

From the observed Zeeman structure in the bosonic isotopes, we can extract the Land\'{e} $\gJ$-factor. This is an important quantity, e.\,g.\, to describe the atomic interaction with an external magnetic field, to describe the interaction between different atoms via their magnetic dipoles, and to benchmark atomic spectrum calculations. Here, we use the relative frequencies of the $\pi$- and $\sigma^+$- transitions to determine the value of the $\gJp$-factor for the excited state. In small magnetic fields, the Zeeman splitting is linear and the transition frequencies can be written as
\begin{equation}
\label{Eq:Pi_Zeeman}
    \nu^\pi = \nu_0 -  m_J ( \gJ - \gJp) \mu_B B / h,
\end{equation}
for the $\pi$-transition and
\begin{equation}
\label{Eq:Sigma_Zeeman}
    \nu^{\sigma +} = \nu_0 - [m_J \gJ  - (m_{J}+1) \gJp] \mu_B B / h,
\end{equation}
for the $\sigma^+$-transition, where $\nu_0$ is the absolute transition frequency at $B = \SI{0}{G}$, $\mu_B$ is the Bohr magneton and $h$ is the Planck constant. By taking the difference of Eq.\,\ref{Eq:Sigma_Zeeman} from Eq.\,\ref{Eq:Pi_Zeeman} one obtains
\begin{equation}
\label{Eq:gJ}
    \gJp = \frac{(\nu^{\sigma +} - \nu^\pi)h}{\mu_B B},
\end{equation}
which allows us to extract the value of $\gJp$, where the uncertainties are arising from the measured frequencies $\nu^\pi$, $\nu^{\sigma +}$ and the applied $B$.

We calibrate $B$, before and after each spectroscopic measurement, by driving the atomic radio-frequency transition of the atoms in \gs~from $m_J = -6 \to m_J = -5$. We evaluate possible drifts of $B$ from these spectroscopy measurements and estimate them to be $\approx \SI{1}{mG}$. This uncertainty on $B$ represents the dominant limitation on the precision of our measurements.

Figure\,\ref{fig:FigMf_andgJ}(b) shows the experimentally extracted values of the $\gJp$-factor as a function of the isotope mass number for the four bosonic isotopes. We find that, as expected, the values for the $\gJp$-factor are the same within one standard deviation for all four isotopes. We combine the results by calculating the weighted mean and determine the $\gJp$-factor of \es~to
\begin{equation}
    \gJp = 1.2599(5).
\end{equation}
The individual $\gJp$-factors are weighted by their standard deviation and the final error corresponds to the combined $1\sigma$-standard deviation.  We compare our experimentally determined $\gJp$-factor to the value specified in the NIST database~\cite{NIST2020}, $\gJp^\mathrm{NIST} = \num{1.266}$, and find agreement at the \SI{1}{\%} level. A careful study of systematic effects such as calibration errors on the magnetic field or collisional shifts (not included in the presented uncertainty) could refine this comparison further, providing a useful benchmark for atomic structure calculations.

\section{Isotope shift}

In addition to the $\gJp$-factor, the high-resolution spectroscopy allows us to extract the isotope shift between the four bosonic isotopes with high precision. Because the $\pi$-transition is less sensitive to magnetic field fluctuations, we fix $B$ and, with the knowledge of the FSR, determine the relative frequency difference directly from the individual transitions. Table~\ref{tab:IsotopesShift} gives the isotope shifts relative to the transition frequency of the $^{168}$Er isotope.

\begin{table}
\caption{Isotope shifts for three bosonic isotopes in dependence of the $^{168}$Er isotope. The error bars denote the statistical error, mainly given by uncertainties of $B$. Systematic errors are not taken into account.}
\label{tab:IsotopesShift}
\begin{ruledtabular}
    \begin{tabular}{cccc}
         & isotope pair & $\nu_0 - \nu_0^{168}$ (MHz) &  \\
        \hline
         & $164-168$  &  -2732.290(3) &  \\
         & $166-168$ & -1371.710(3)  & \\
        & $170-168$ & 1414.920(5) &  \\
    \end{tabular}
\end{ruledtabular}
\end{table}

At leading order, isotope shifts are caused by two effects, the field shift and the mass shift, which arise from the change of the nuclear size and the mass, respectively. Here, for the involved $4f \rightarrow 5d$ transition, both of the contributions are comparably large~\cite{Budker2008,Lipert1993}. Isotope shifts of two different transitions, plotted against each other, follow at leading order a linear dependence which is referred to as King's linearity~\cite{King1963}. Violations of the linearity can provide an exceptional insight into intra-nuclear interactions and can help to shed light on processes that are not described by the current theory of the standard model~\cite{Delaunay2017,Mikami2017,Berengut2018,Flambaum2018,Miyake2019,Counts2020}. The availability of four isotopes that have zero nuclear spin and a large number of different narrow transitions make erbium a potential candidate for investigations along this route. In particular, erbium features further narrow transitions that involve the excitation of an electron from the $4f$ orbital to the $5d$ orbital. Consequently, in combination with our transition at \SI{1299}{nm}, the non-linearity might be less sensitive to field shifts induced by different electron configurations and therefore interesting for future investigations~\cite{Mikami2017,Delaunay2017,Flambaum2018}.

\section{Coherent control and lifetime measurements}

An important opportunity that comes along with narrow line transitions and plays a fundamental role, e.\,g.\, in quantum information and communication protocols, is the possibility to coherently control the atomic state. To demonstrate the ability to drive coherent excitations, we measure Rabi oscillations on the closed $\sigma^-$-transition for a thermal cloud of the $^{168}$Er isotope. We use a thermal cloud to reduce the effect of interactions, by allowing for lower densities. From theoretical calculations (see description in Sec.\,\ref{Sec:Theory}), it is expected that the atomic polarizability of atoms in \es~is very low, or even negative at \SI{1064}{nm}, depending on the polarization of the trapping light. Therefore, we transfer the atoms after evaporation into a crossed ODT that is created by two intersecting laser beams at \SI{532.2}{nm} and \SI{1570}{nm}, resulting in trap frequencies of $(\omega_x, \omega_y, \omega_z) = 2\pi [232(6),117(7),209(3)]\,$Hz for the ground-state atoms. At this stage, we measure $\num{2e4}$ atoms at a temperature of $T\approx\SI{700}{nK}$, corresponding to a peak density of about \SI{9e12}{cm^{-3}}. After the preparation of the atomic cloud, we shine a resonant narrow-line laser with a peak intensity of about \SI{4.9}{W\per cm^2} for a pulse duration of $t_\mathrm{pulse}$ onto the atomic sample and measure the atom number in \gs~after a time of flight of \SI{10}{ms}.

\begin{figure}
    \centering
    \includegraphics[width = 0.48\textwidth]{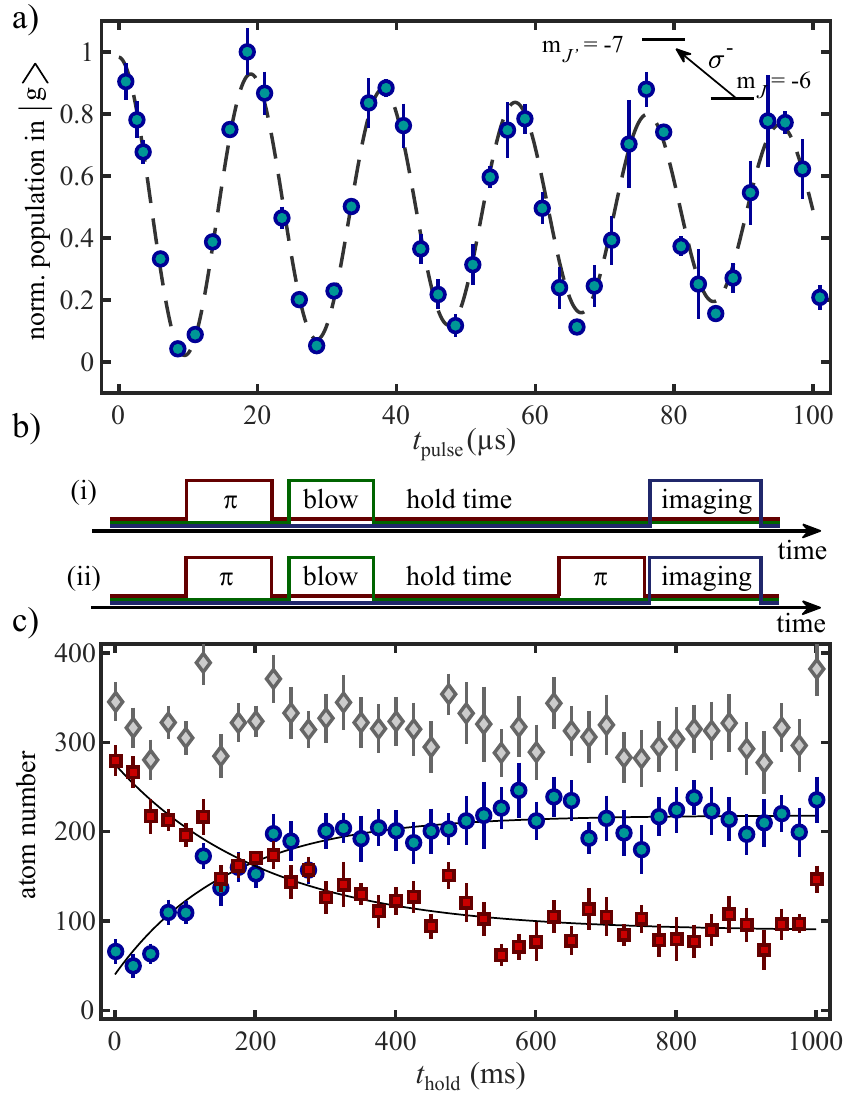}
    \caption{(a) Coherent Rabi oscillations for $^{168}$Er on the $\ket{m_J = -6} \to \ket{m_{J'} = -7}$ transition. Shown is the normalized atom number in \gs~against the duration of the laser pulse. The dashed line represents a fit of a damped sinusoidal oscillation to the experimental data points. (b) Schematic illustration of the two applied measurement sequences to extract state populations~(see main text). (c) blue circles (red squares) represent the atom number in \gs~(\es) in dependence of the hold time for the measurements sequence (i) ((ii)). The grey diamonds represent the total atom number. The solid lines are exponential fits to the experimental data. Error bars denote the standard error of 4 (a) and 10 (c) repetitions.}
    \label{fig:Fig3}
\end{figure}

Figure~\ref{fig:Fig3}(a) shows the population in \gs, normalized to the maximum atom number as a function of $t_\mathrm{pulse}$. We observe a damped oscillation of the population in \gs, which is well described via $p_g(t) = 0.5 e^{\nicefrac{-t}{\tau_\mathrm{osc}}}\cos ( \Omega _R t ) + 0.5$, where $\Omega _R$ is the Rabi frequency and $\tau_\mathrm{osc}$ is the $\nicefrac{1}{e}$ decay time of the contrast of the oscillation. We find that $\Omega _R = 2\pi \times \SI{50.02(8)}{kHz}$, which corresponds to a normalized Rabi frequency of $\Omega _R^\mathrm{norm.} = 2\pi \times \SI{0.34(3)}{Hz \per \sqrt{mW \per cm^2}}$. For the decay time of the contrast we find $\tau_\mathrm{osc} = \SI{192(20)}{\micro s}$, indicating a strong decoherence mechanism. Possible mechanisms that might lead to the decoherence are, e.\,g.\, atomic interactions, intensity noise on the trapping light, intensity inhomogeneities of the probe light over the atomic cloud, or fluctuations of the magnetic field. Nonetheless, the coherent control allows us transfer atoms from \gs~to \es~ with an efficiency $>\SI{97}{\%}$.

The ability to transfer atoms from \gs~to \es~with high efficiency enables us to measure the lifetime of atoms in \es. At high densities we observe a short lifetime of the sample in \es, suggesting a density-dependent loss mechanism, similarly to Ref.\,\cite{Bishof2011}. Therefore, we reduce the atom number to $N \approx 300$ atoms by using a shorter loading time of the magneto-optical trap and we stop the evaporative cooling process at an earlier stage, leading to a temperature of $T\approx \SI{1}{\micro K}$. Here, we obtain a peak number density of about $\SI{7.5e10}{cm^{-3}}$.

To measure the lifetime of \es, we carry out two complementary measurements. First, we perform a $\pi$-pulse to transfer \gs~atoms to \es~with high efficiency.
To obtain a pure sample of \es~atoms, we remove the small remaining fraction of \gs~atoms with a resonant pulse at \SI{401}{nm}. Note that, the light at \SI{401}{nm} is not resonant for atoms in \es. We then measure the lifetime of the excited sample in two independent measurements, (i) the number of atoms in \gs, which decayed from \es~due to spontaneous emission, and (ii) the atomic population in \es~by applying a second $\pi$-pulse to invert the populations in \es~and \gs~in order to directly measure the excited-state atoms; see Fig.\,\ref{fig:Fig3}(b).

Figure~\ref{fig:Fig3}(c) shows the measured atom number for both measurement sequences at different holding times $t_\mathrm{hold}$. We observe a decay of the atom number in \es, which is consistent with the simultaneous growth of the atom number in \gs. Note that, the sum of the atom number in both states remains constant over the observed timescale, indicating that atoms in \es, indeed, decay dominantly to \gs. We extract the lifetime, for both measurement protocols, by fitting an exponential function $N(t) = a e^{-t/\tau _e} + d$ to the non-averaged atom numbers. Here, $a$ denotes the amplitude and $d$ the offset of the growth (decay) of the atom number in \gs~(\es). The characteristic time $\tau _e$ represents the lifetime of \es. We extract a lifetime of \SI{162(23)}{ms} [\SI{212(33)}{ms}] through the measurement sequence (i) [(ii)]. We combine both results by calculating the weighted mean and obtain a mean lifetime of
\begin{equation}
   \tau _e = \SI{178(19)}{ms}.
\end{equation}
This lifetime corresponds to a natural linewidth of \SI{0.9(1)}{Hz}, which is in agreement within error bars with the theoretical predicted value of \SI{2(1)}{Hz} in Ref.\,\cite{Ban2005}. Note that, the measured lifetime is consistent with the natural linewidth predicted from $\Omega _R$ determined above.

\section{Theoretical Predictions}
\label{Sec:Theory}

We compare our experimental findings with the results of a semi-empirical model, which has previously been very successful in predicting the properties of broader optical transitions in erbium and dysprosium~\cite{lepers2016,li2017}.
Our calculations are based on the semi-empirical method provided by the COWAN suite of codes \cite{cowan1981, kramida2019}, and extended by us \cite{lepers2016}. In a first step, \textit{ab initio} radial wave functions $P_{n\ell}$ for all the subshells $n\ell$ of the considered configurations, with $n$ and $\ell$ being the principal and orbital quantum numbers, are computed with the relativistic Hartree-Fock (HFR) method. Those wave functions are then used to calculate energy parameters that are the building blocks of the atomic Hamiltonian. In a second step, the energy parameters are adjusted so that the eigenvalues of the Hamiltonian best fit the experimental energies of the NIST database~\cite{NIST2020}, using Kramida's version of the least-square fitting COWAN code RCE \cite{kramida2019}. The $P_{n\ell}$ wave functions also serve to calculate the mono-electronic transition integrals $\langle n\ell |r| n'\ell' \rangle = \int dr P_{n\ell}(r) r P_{n'\ell'}(r)$, that are the building blocks of Einstein coefficients for spontaneous emission $A_{ik}$. In a third step, the $\langle n\ell |r| n'\ell' \rangle$ integrals are adjusted to minimize the difference between experimental and theoretical $A_{ik}$ coefficients \cite{lepers2016}.

For the even-parity levels of erbium, we use the same energy parameters as Ref.\,\cite{Becher2018}. Briefly, the even electronic configurations are separated into three groups:
\begin{gather*}
        4f^{12}6s^{2} + 4f^{12}5d6s + 4f^{11}6s^{2}6p, \\
        4f^{12}6s^{2} + 4f^{11}5d6s6p,
\end{gather*}
and
\begin{gather*}
        4f^{12}6s^{2} + 4f^{12}6s7s + 4f^{12}6s6d + 4f^{12}6p^{2}.
\end{gather*}
Each group is associated with a different least-square fitting calculation with experimental levels belonging to the corresponding configurations.

Compared to Refs.\,\cite{Lepers2014,Becher2018}, the odd-parity level calculations have been improved by adding some high-lying experimental energy levels that were previously excluded from the fitting procedure, as well as incorporating a larger number of free configuration-interaction parameters into the fitting procedure.  The following configurations are included in the calculation: $4f^{11} 5d 6s^2$, $4f^{11} 5d^2 6s$, $4f^{12} 6s 6p$, $4f^{12} 5d 6p$ and $4f^{13} 6s$. The latter is included for technical purpose, but does not play a physical role. The fitting procedure is performed using a total of 30 free groups of parameters and 219 levels. The standard deviation between experimental and calculated energies is equal to \SI{53}{cm^{-1}}, which is satisfactory for a semi-empirical calculation. Details on the parameters for the first four odd parity configurations are given in Ref.\,\cite{supmat}.

The $\langle n\ell |r| n'\ell' \rangle$ transition integrals were adjusted using the set of experimental $A_{ik}$ coefficients of Ref.~\cite{lawler2010}, especially the transitions involving levels of the ground-state configuration [Xe]$4f^{12} 6s^2$. Following Ref.\,\cite{lepers2016}, we seek to minimize the standard deviation $\sigma_A$ on Einstein coefficients $A_{ik}$ \cite{lepers2016}. Because the latter is poorly sensitive to $\langle 4f |r| 5d \rangle$, we could not find a value of that integral minimizing $\sigma_A$, we have taken a scaling factor with respect to the HFR integral equal to $f_{4f,5d} = 0.95$, following previous works on dysprosium \cite{li2017} and holmium \cite{li2017a}. We applied the fitting procedure on $\langle 6s |r| 6p \rangle$, and found $f_{6s,6p} = 0.786$. We have fitted 77 experimental lines, and found a standard deviation $\sigma_A = 8.085 \times 10^6$~s$^{-1}$.

With this optimized set of energies and transition integrals, we have calculated the polarizabilities of the ground and excited states using the sum-over-state formula coming from second-order perturbation theory. The polarizability of the excited level also depends on $\langle 5d |r| 6p \rangle$, for which we took a scaling factor of 0.8.

\begin{figure*}
    \centering
    \includegraphics[width =1\textwidth]{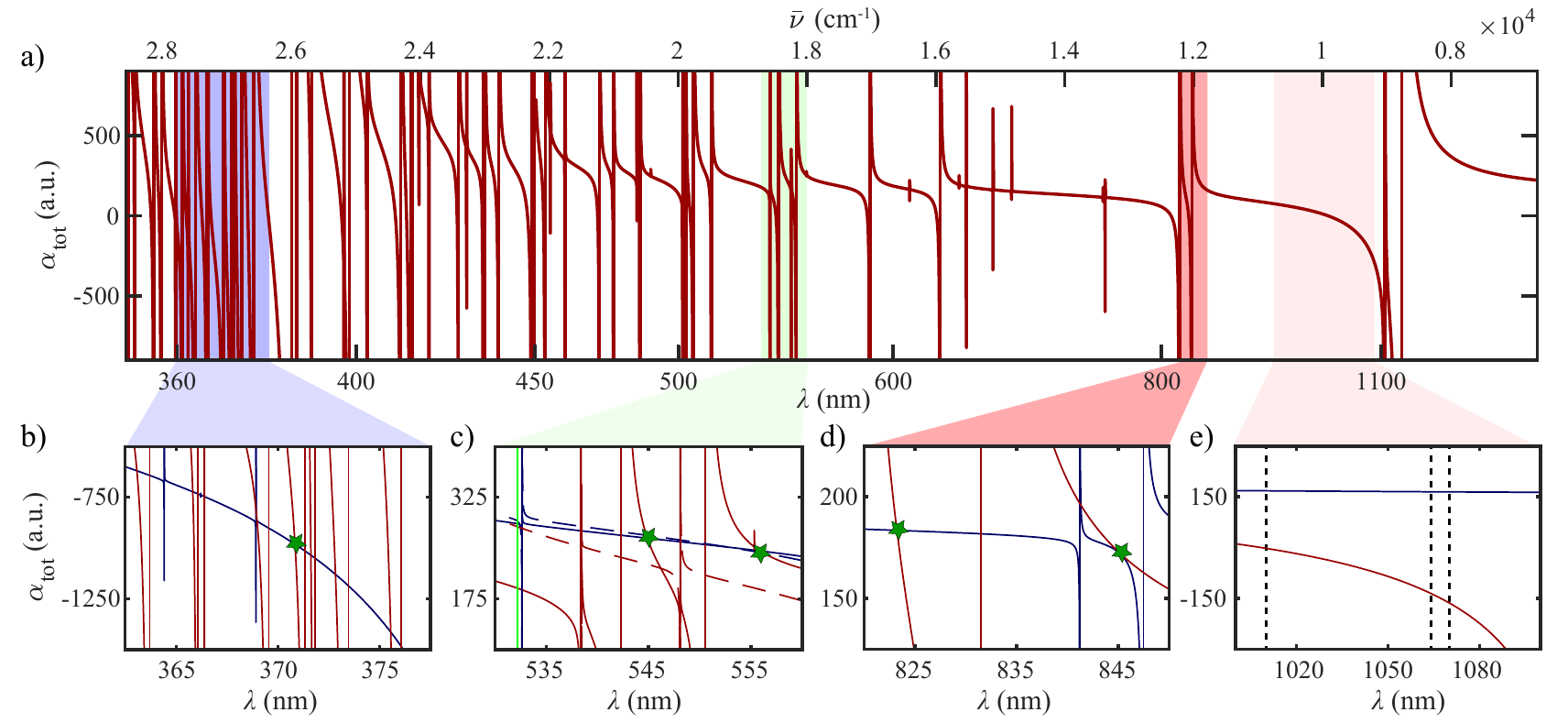}
   \caption{(a) Theoretically calculated $\alpha(\omega)$ for \es~as a function of the wavelength $\lambda$ for $\pi$-polarized light. (b)-(e) Zoom-in into specific wavelength regions showing $\alpha(\omega)$ for \es~(red) and \gs~(blue). The green stars indicate possible magic wavelengths. In (c), the dashed line represents $\alpha(\omega)$ for $\sigma^-$-polarized light. The green solid line indicates the wavelength for the measurement in Fig.\,\ref{fig:PolarizabilityRatio}. The dashed black lines in (e) denote the wavelength of \SI{1010}{nm}, \SI{1064}{nm}, and \SI{1070}{nm}.}
   \label{fig:Fig_Polarizability}
\end{figure*}

From our theory we obtain a wavenumber of $\Bar{\nu}^\mathrm{th} = \SI{7729.3}{cm^{-1}}$, a $\gJp$-factor of $\gJp^\mathrm{th} = \num{1.2622}$, and an excited-state lifetime of $\tau^\mathrm{th} = \SI{602}{ms}$. For $\Bar{\nu}^\mathrm{th}$ and $\gJp^\mathrm{th}$ we find satisfactory agreement with the values reported from the current experimental work. Note that, by including the experimental energy parameters in the theoretical calculations (compared with Ref.\cite{Allehabi2020}), we obtain better agreement with the experimental data. The extracted lifetime is about a factor of $3$ longer compared to the experimentally measured value. This discrepancy comes from the fact that the underlying transition dipole moment involves small components in the eigenvector associated with level $|e\rangle$. Those small components are more difficult to optimize, as they are less affected by the least-square fitting procedure on energies.

Figure~\ref{fig:Fig_Polarizability}(a) shows the calculated atomic polarizability $\alpha(\omega)$ for \es~in a broad wavelength range from \SI{350}{nm} to \SI{1500}{nm}. The polarizability spectrum becomes very dense at lower wavelengths. The background value of $\alpha (\omega)$ is dominantly positive, however, a strong transition at around \SI{1140}{nm} causes a negative value of $\alpha (\omega)$ around \SI{1064}{nm}, commonly used for optical dipole traps. Further, this strong transition creates two interesting situations appearing at \SI{1010}{nm} and \SI{1070}{nm}. Here, while $\alpha(\omega)$ is finite for \gs, $\alpha(\omega)$ of \es~is either $0$ (\SI{1010}{nm}) or has the same absolute value with opposite sign (\SI{1070}{nm}). These circumstances are beneficial for the realization of spin-dependent lattice configurations~\cite{Yang2017}. Due to the weak coupling to \gs, the effect of the transition at \SI{1299}{nm} is not visible in this plotting range.

\section{Magic-wavelength conditions}

A very important ingredient for the coherent control of our two-level system is the atomic polarizability of each state, $\alpha (\omega)$, and their ratio. To extend the coherence time, it is important to work at a magic-wavelength condition, because a different $\alpha (\omega)$ for \gs~and \es~constitutes a source for dephasing and leads therefore to a reduced coherence time.
For the ground state, $\alpha (\omega)$ has been extensively studied and measured at various wavelengths~\cite{lepers2016,Becher2018}. Contrary, prior to this work, the polarizability of the excited state was not known.

The large number of optical transitions in atomic erbium provides many possibilities for magic-wavelength conditions. We use our theoretical calculations to identify several interesting crossings of the atomic polarizability of the ground state and the excited state. Figure~\ref{fig:Fig_Polarizability}(b)-(d) show $\alpha (\omega)$ of \gs~and \es~for $\pi$-polarized light in three different wavelength ranges, all containing crossings which are promising for the realization of optical potentials at magic-wavelength conditions. Note that for both states, the atoms are considered to be in the lowest Zeeman level. The negative values of $\alpha (\omega)$ for both, ground state and excited state, at $\approx \SI{370}{nm}$ potentially allows for the realization of blue-detuned optical lattices. The crossings of the polarizabilities at \SI{545}{nm}, \SI{557}{nm}, \SI{822}{nm}, and \SI{845}{nm} are positive, and therefore red-detuned potentials can be created.

Moreover, in erbium, as in other lanthanides, the total polarizability is anisotropic with important vectorial and tensorial contributions. This allows for instance an orientation tuning of the optical dipole potential~\cite{Kao2017,Chalopin2018,Becher2018}. In detail, the atomic polarizability $\alpha (\omega)$ of an atom with non-zero angular-momentum $\mathbf{J}$, that is placed in an external laser field oscillating at a frequency $\omega$, can be expressed as:
\begin{eqnarray}
        \alpha (\omega) &=& \alpha_s (\omega) + i \frac{[\mathbf{u^*} \times \mathbf{u}] \cdot \mathbf{J}}{2J} \alpha_v (\omega) \nonumber\\
&& + \frac{J(J+1) -3m_j^2}{J(2J-1)} \frac{1- 3 \cos^2 \theta_p}{2} \alpha_t (\omega),
\label{Eq:alpha}
\end{eqnarray}
where $\alpha_s (\omega)$, $\alpha_v (\omega)$, and $\alpha_t (\omega)$ are the scalar, the vectorial, and the tensorial polarizabilities and $\mathbf{u}$ is the polarization vector of the laser field. The angle $\theta_p$ defines the orientation of $\mathbf{u}$ with respect to $\mathbf{B}$. Importantly, the dependence of $\alpha (\omega)$ on $\mathbf{u}$ and $\theta_p$ allows to tune $\alpha (\omega)$ for a fixed $\omega$. This additional tunability on the polarization of the trapping light increases the possibilities for magic-trapping conditions~\cite{Kim2013,Chalopin2018}. In Figure~\ref{fig:Fig_Polarizability}(c) we plot $\alpha (\omega)$ of \gs~and \es~for both, $\pi$- and $\sigma^-$-polarization. While for $\pi$-polarization the polarizability of the two states is different, we observe a crossing around \SI{532}{nm} for $\sigma^-$-polarization.

\begin{figure}
    \centering
    \includegraphics[width =0.48\textwidth]{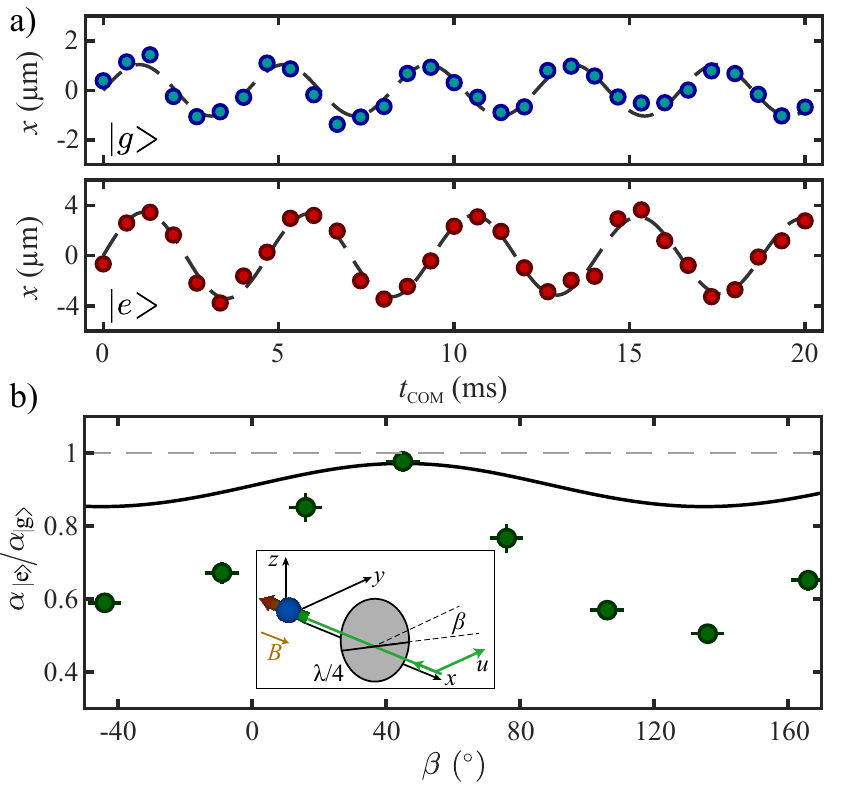}
    \caption{(a) Center position of the thermal cloud in dependence of the hold time for \gs~(blue) and \es~(red) for the case of linearly polarized light with $\mathbf{u}$ pointing along $y$ and $\mathbf{B}$ aligned opposite to the propagation axis of the \SI{532.2}{nm} trapping light. (b) Polarizability ratio $\nicefrac{\alpha_{\ket{e}}}{\alpha_{\ket{g}}}$ for different polarizations of the trapping light defined by the angle $\beta$ between $\mathbf{u}$ and the optical axis of the $\nicefrac{\lambda}{4}$-waveplate. The black solid line represents theoretical results obtained from a sum-over-states method. The inset illustrates the experimental configuration for the shown data.}
    \label{fig:PolarizabilityRatio}
\end{figure}

The determination of the absolute value of the atomic polarizability is challenging since possible methods rely on accurate knowledge of the waist of the laser beam. To circumvent this issue, we measure the ratio of the trapping frequency~\cite{Ravensbergen2018} of \gs~and \es, which is independent of the exact parameters of the trapping beam. The atomic polarizability of \gs~and \es~are related to the trap frequencies via $\nicefrac{\alpha_{\ket{e}}}{\alpha_{\ket{g}}} = \left( \nicefrac{\omega_{\ket{e}}}{\omega_{\ket{g}}} \right)^2$.
We use a thermal cloud of $N \approx\, \num{2e3}$ atoms at $T\approx\SI{1}{\micro K}$ and hold in the same optical trap as for the previous measurement of the lifetime. We orientate $B$ anti-parallel to the propagation of the light at \SI{532.2}{nm} and we control the light polarization of the latter by using a $\nicefrac{\lambda}{4}$-waveplate; see inset Fig.~\ref{fig:PolarizabilityRatio}(b).

First, we focus on the case where the trapping light is horizontally polarized along the direction of $y$.
We measure the trap frequency along the vertical direction by exciting a center-of-mass (COM) mode, which is dominantly contributed by the horizontal trapping beam at \SI{532.2}{nm}. The measurement of the COM for \es~is performed similarly to the protocol (ii) in Fig.\,\ref{fig:Fig3}(b) and we excite the COM after the blow-pulse. Figure~\ref{fig:PolarizabilityRatio}(a) exemplary shows the evolution of the center position of the atomic cloud for variable hold times $t_\mathrm{COM}$. For both atomic states, we observe several periods of the oscillation of the center position and the oscillation frequency is extracted by fitting the data with a damped sinusoidal function. The relative polarizability for a linear polarized dipole trap is obtained to be $\nicefrac{\alpha_{\ket{e}}}{\alpha_{\ket{g}}} = \num{0.77(1)}$.

We now probe the anisotropy of $\alpha (\omega)$ by repeating the trap frequency measurements for different polarization compositions of the trapping light~\cite{PolarizationControl}, as shown in Figure~\ref{fig:PolarizabilityRatio}(b). The light polarization is denoted by the angle $\beta$. The ratio $\nicefrac{\alpha_{\ket{e}}}{\alpha_{\ket{g}}}$ has a large variation and reaches a maximum of $\num{0.98(3)}$ for $\sigma^-$-polarized light ($\beta\approx\SI{45}{\degree}$), which indicates a magic-wavelength condition. Interestingly, even though the theoretical prediction agrees well with the experimental results at \SI{45}{\degree} and qualitatively follows the data, we observe an increased deviation for the other polarization compositions, which indicates an underestimation of the vectorial polarizability.

\section{Conclusion}

In conclusion, we have observed the narrow inner-shell orbital transition at \SI{1299}{nm} for the four bosonic, as well as for the fermionic isotope. We characterized the transition by measuring the $\gJp$-factor and the atomic polarizability at \SI{532.2}{nm}, which we compare to a semi-empirical model. Further, we demonstrated the ability to coherently control the atomic state. The narrow transition, at a wavelength within the telecom window, with a linewidth of \SI{0.9(1)}{Hz} and a related long lifetime of \SI{178(19)}{ms}, represents a very versatile tool with outstanding potential for a broad range of possible applications.

For the realization of efficient long-distance quantum communication, atom-photon interfaces at wavelengths that are compatible with telecom wavelengths are highly desired~\cite{Kimble2008,Reiserer2015,Wehner2018}. To date, most of the existing approaches rely either on frequency conversion of the photons~\cite{Dreau2018,Krutyanskiy2019}, leading to undesired noise and reduced efficiency, or suffer from various other dephasing mechanisms~\cite{Pfaff2014,Bauch2020}. Here, our transition at \SI{1299}{nm}, lying within the O-band of the telecom wavelength architecture, is a potential candidate to circumvent these issues. Finally, the photon storage time can be drastically improved by collective scattering in ordered atomic arrays, exceeding the natural lifetime~\cite{Shahmoon2017,Asenjo2017}.

The relatively long wavelength of this transition may also provide a favorable platform for studies of collective scattering from ordered atomic samples. Such effects, which include geometry-dependent enhancement or suppression of emission~\cite{Facchinetti2016,Rui2020}, can be more significant when the spacing between atoms is small relative to the transition wavelength.  For our \SI{266}{nm} spacing, typical of lattice confinement with \SI{532}{nm} light~\cite{Baier2016}, this condition is well met in our system, in contrast to the more common situation present in alkaline atoms where the wavelength of trapping light typically exceeds the wavelength of the scattering transition.

Additionally, the advantages of encoding qubits in the ground and the metastable state and the possibility for independent control of atomic motions by lattice light is promising for quantum computational tasks~\cite{Daley2008,Gorshkov2009,Heinz2020}.
Moreover, the contact interaction can be tuned using the technique of a narrow-line optical Feshbach resonance, where the system suffers only weak atom loss~\cite{Fatemi2000,Theis2004,Blatt2011}.

Finally, this transition enables the coherent control of magnetic Zeeman levels for dipolar bosonic atoms, which has been elusive so far, due to the absent hyperfine interactions~\cite{Baier2018}. The 13 magnetic Zeeman levels in the ground state of erbium can be employed as spin states and allow for the realization of large bosonic spin systems with dipolar interactions. In combination with optical lattices, this enables the possibility to study many-body dynamics in extended Bose-Hubbard models~\cite{Lahaye2009,Baranov2012}. Generally, in such large spin systems, the long lifetime of the excited state is helpful for the realization of advanced, spin-resolved imaging-shelving techniques~\cite{Staanum2004,Yang2020}.

\begin{acknowledgments}

We thank H.\, Ritsch and M.\,A.\,Norcia for insightful discussions and M.\,A.\,Norcia for careful reading of the manuscript. This work is financially supported through an ERC Consolidator Grant (RARE, no.\,681432), a DFG/FWF (FOR 2247/I4317-N36) and a joint-project grant from the FWF (I4426, RSF/Russland 2019). L.\,C.\,acknowledges the support of the FWF via the Elise Richter Fellowship number V792. We also acknowledge the Innsbruck Laser Core Facility, financed by the Austrian Federal Ministry of Science, Research and Economy. G.\,H.\,and M.\,L.\,acknowledge support from the NeoDip project (ANR-19-CE30-0018-01 from ``Agence Nationale de la Recherche''). M.\,L.\,also acknowledges the financial support of {}``R{\'e}gion Bourgogne Franche Comt{\'e}'' under the projet 2018Y.07063 {}``Th{\'e}CUP''. Calculations have been performed using HPC resources from DNUM CCUB (Centre de Calcul de l'Universit\'e de Bourgogne).

\end{acknowledgments}

* Correspondence and requests for materials
should be addressed to Francesca.Ferlaino@uibk.ac.at.


\begin{thebibliography}{81}
\bibitem [{Wav()}]{WavemeterCalib}%
	\BibitemOpen
	\href@noop {} {}\bibinfo {note} {The wavelength meter
		(\r{A}ngstrom/HighFinesse Ws/7R) is calibrated on the transition at
		\SI{729}{nm} of the $^{40}$Ca$^+$ ion.}\BibitemShut {Stop}%
\bibitem [{Pol()}]{PolarizationControl}%
	\BibitemOpen
	\href@noop {} {}\bibinfo {note} {We control the light polarization by
		changing the angle $\beta$ of the $\nicefrac{\lambda}{4}$-waveplate, which
		defines the angle between the incoming horizontal polarization and the fast
		axis of the waveplate. $\mathbf{u} =
		(\cos(\beta),0,i\sin(\beta))$.}\BibitemShut {Stop}%
\end{thebibliography}

%


\clearpage

\renewcommand\thetable{S\arabic{table}}
\setcounter{table}{0}

\onecolumngrid
\vspace*{0.5cm}
\begin{center}
    \textbf{SUPPLEMENTARY MATERIALS}
\end{center}
\vspace*{0.5cm}
\twocolumngrid

\section*{Odd-parity energy parameters}

Here we present the parameters of the first four odd-parity configurations of neutral erbium. The ones of the $4f^{13} 6s$ configuration are not shown, as the latter does not play any physical role in the level interpretation.

Tables~\ref{tab:Er_param_1} and~\ref{tab:Er_param_2} show the one-configuration parameters, like the direct $F^k$, exchange $G^k$, or spin-orbit $\zeta_{n\ell}$ integrals. Table~\ref{tab:Er_param_3} shows the configuration-interaction ones. We also give the scaling factor $f_X = X_\mathrm{fit} / X_\mathrm{HFR}$ between the fitted and \textit{ab initio} value of the parameter $X$. During the fitting procedure, some groups of parameters are constrained to vary within the same scaling factors; such groups are characterized by the same $f_X = r_n$ value in the second column of Tables \ref{tab:Er_param_1}--\ref{tab:Er_param_3}. The word ``fix" means that the corresponding parameters are not adjusted. Finally, we use so-called ``effective" parameters, like $\alpha$, $\beta$, $\gamma$ or $F_1(4f5d)$, which cannot be calculated \textit{ab initio}, but which are there to compensate the absence of electronic configurations not included in the model. Their initial values are known from previous studies.

\onecolumngrid
\vspace{\columnsep}


\begin{table}[!h]
\begin{minipage}{\textwidth}
\caption{\label{tab:Er_param_1} Parameter names, constraints (see text), fitted values and scaling factors $f_X = X_\mathrm{fit} / X_\mathrm{HFR}$, for the $4f^{11} 5d 6s^2$ and $4f^{11} 5d^2 6s$ configurations of neutral Er. All parameters are in cm$^{-1}$.}
\begin{ruledtabular}
\begin{tabular}{ccrrrr}
Parameter $X$ & Constraint & $X_\mathrm{fit}$ & $f_X$ & $X_\mathrm{fit}$ & $f_X$ \\
\hline \\
 & & \multicolumn{2}{c}{4f$^{11}$  5d  6s$^2$} & \multicolumn{2}{c}{4f$^{11}$  5d$^2$  6s} \\
\cline{3-4} \cline{5-6} 
 & & & & &   \\
$E_\mathrm{av}$ &    & 46412.0  &     & 65531.6 & \\
$F^2$(4f 4f) & $r_1$ & 98177.6 & 0.761 & 98004.7 & 0.761\\
$F^4$(4f 4f) & $r_2$ & 69264.0   & 0.856 & 69134.0 & 0.856 \\
$F^6$(4f 4f) & $r_3$ & 50068.0   & 0.861 & 49972.2 & 0.861 \\
$\alpha$     & $r_4$ &    20.0 &       &    20.0 &  \\
$\beta$      & fix   &  -650.0 &       &  -650.0 &  \\
$\gamma$     & fix   &  2000.0 &       &  2000.0 &  \\
$F^2$(5d 5d) &       &         &       & 21668.3 & 0.663\\
$F^4$(5d 5d) &       &         &       & 17208.2 & 0.831\\
$\zeta_{4f}$ & $r_5$ &  2389.8 & 0.984 &  2387.8 & 0.984\\
$\zeta_{5d}$ & $r_6$ &   788.2 & 0.831 &   652.6 & 0.831\\
$F^1$(4f 5d) & $r_7$ &   741.7 &       &   741.7 & \\
$F^2$(4f 5d) & $r_8$ & 15711.7 & 0.775 & 13597.8 & 0.775\\
$F^4$(4f 5d) & $r_9$ & 10558.2 & 1.149 &  8970.9 & 1.149\\
$G^1$(4f 5d) & $r_{10}$ & 5054.1 & 0.580 &  4325.8 & 0.580\\
$G^2$(4f 5d) & $r_{11}$ & 1717.4 &       &  1717.4 & \\
$G^3$(4f 5d) & $r_{12}$ & 6400.3 & 0.928 &  5422.6 & 0.928\\
$G^4$(4f 5d) & $r_{13}$ & 1630.6 &       &  1630.6 & \\
$G^5$(4f 5d) & $r_{14}$ & 3809.7 & 0.732 &  3216.7 & 0.732\\
$G^3$(4f 6s) & $r_{15}$ &        &       &  1254.3 & 0.844\\
$G^2$(5d 6s) & $r_{17}$ &        &       & 11696.8 & 0.609\\
\end{tabular}
\end{ruledtabular}
\end{minipage}
\end{table}


\vspace{\columnsep}
\twocolumngrid


\begin{table*}
\caption{\label{tab:Er_param_2}  Parameter names, constraints (see text), fitted values and scaling factors $f_X = X_\mathrm{fit} / X_\mathrm{HFR}$, for the $4f^{12} 6s 6p$ and $4f^{12} 5d 6p$ configurations of neutral Er. All parameters are in cm$^{-1}$.}
\begin{ruledtabular}
\begin{tabular}{ccrrrr}
Parameter $X$ & Constraint & $X_\mathrm{fit}$ & $f_X$  & $X_\mathrm{fit}$ & $f_X$ \\
\hline \\
& & \multicolumn{2}{c}{4f$^{12}$  6s  6p} &
\multicolumn{2}{c}{4f$^{12}$  5d  6p} \\
\cline{3-4} \cline{5-6} \\
$E_\mathrm{av}$ &  & 36278.6 &  & 60020.5 & \\
$F^2$(4f 4f) & $r_1$ & 92716.3 & 0.761 & 92496.8 & 0.761\\
$F^4$(4f 4f) & $r_2$ & 65123.3 & 0.856 & 64956.1 & 0.856 \\
$F^6$(4f 4f) & $r_3$ & 46996.3 & 0.861 & 46876.2 & 0.861 \\
$\alpha$ & $r_4$ & 20.0 &  & 20.0 &  \\
$\beta$ & fix & -650.0 &  & -650.0 &  \\
$\gamma$ & fix & 2000.0 &  & 2000.0 &  \\
$\zeta_{4f}$ & $r_5$ & 2250.3 & 0.984 & 2250.3 & 0.985\\
$\zeta_{5d}$ & $r_6$ &  &  & 454.7 & 0.831\\
$\zeta_{6p}$ & $r_{18}$ & 1513.0 & 1.461 & 1119.6 & 1.461\\
$F^1$(4f 5d) & $r_7$ &  &  & 741.7 & \\
$F^2$(4f 5d) & $r_8$ &  &  & 11092.6 & 0.775\\
$F^4$(4f 5d) & $r_9$ &  &  & 7222.8 & 1.149\\
$F^1$(4f 6p) & fix & 100.0 &  & 150.0 & \\
$F^2$(4f 6p) & $r_{19}$ & 3776.9 & 1.156 & 3006.1 & 1.151 \\
$F^2$(6p 5d) & fix &  &  & 11470.1 & 0.794 \\
$G^1$(4f 5d) & fix &  &  & 3897.8 & 0.585\\
$G^2$(4f 5d) & fix &  &  & 1091.6 & \\
$G^3$(4f 5d) & fix &  &  & 4397.4 & 0.895\\
$G^4$(4f 5d) & fix &  &  & 1028.3 & \\
$G^5$(4f 5d) & fix &  &  & 2761.2 & 0.761\\
$G^3$(4f 6s) & $r_{15}$ & 1405.1  & 0.843 &  & \\
$G^2$(4f 6p) & $r_{19}$ & 864.4 & 1.155 & 656.3 & 1.155 \\
$G^4$(4f 6p) & $r_{19}$ & 751.9 & 1.156 & 568.2 & 1.157 \\
$G^1$(6s 6p) &  & 12209.7 & 0.522 &  &  \\
$G^1$(6p 5d) & fix &  &  & 7880.0 & 0.600  \\
$G^3$(6p 5d) & fix &  &  & 5052.0 & 0.600  \\
\end{tabular}
\end{ruledtabular}
\end{table*}



\begin{table}
\caption{\label{tab:Er_param_3} Configuration-interaction parameters: parameter names, constraints (see text), fitted values and scaling factors $f_X = X_\mathrm{fit} / X_\mathrm{HFR}$, for odd-parity configuration pairs of neutral Er. All parameters are in cm$^{-1}$.}
\begin{ruledtabular}
\begin{tabular}{ccrr}
Parameter $X$ & Constraint & $X_\mathrm{fit}$ & $f_X$ \\
\hline \\
 & & \multicolumn{2}{c}{4f$^{11}$  5d  6s$^2$ - 4f$^{11}$  5d$^2$  6s}  \\
\cline{3-4}
 & & &   \\
$R^2$ (4f 6s, 4f 5d) & $r_{21}$ & -799.8 & 0.852 \\
$R^3$ (4f 6s, 4f 5d) & $r_{21}$ & 1128.5 & 1.465 \\
$R^2$ (5d 6s, 5d 5d) & $r_{17}$ & -13663.3 & 0.621 \\
\\
& & \multicolumn{2}{c}{4f$^{11}$  5d  6s$^2$ - 4f$^{12}$  6s  6p} \\
\cline{3-4}
 & & &   \\
$R^1$ (5d 6s, 4f 6p) & $r_{20}$ & -3173.6 & 0.461  \\
$R^3$ (5d 6s, 6p 4f) & $r_{20}$ & -679.9 & 0.461 \\
\\
& & \multicolumn{2}{c}{4f$^{11}$  5d$^2$  6s - 4f$^{12}$  6s  6p} \\
\cline{3-4}
 & & &  \\
$R^1$ (5d 5d, 4f 6p) & $r_{22}$ & 2405.6 & 0.647 \\
$R^3$ (5d 5d, 4f 6p) & $r_{22}$ & 643.7 & 0.647 \\
\\
& & \multicolumn{2}{c}{4f$^{11}$  5d$^2$  6s - 4f$^{12}$  5d  6p} \\
\cline{3-4}
 & & &   \\
$R^1$ (5d 6s, 4f 6p) & $r_{22}$ & -2536.8 & 0.427 \\
$R^3$ (5d 6s, 4f 6p) & $r_{22}$ & -564.5 & 0.427 \\
\\
& & \multicolumn{2}{c}{4f$^{12}$  6s  6p - 4f$^{12}$  5d  6p} \\
\cline{3-4}
 & & &  \\
$R^2$ (4f 6s, 4f 5d) & $r_{23}$ & -5359.9 &  \\
$R^3$ (4f 6s, 5d 4f) & $r_{23}$ & 2758.3 &  \\
$R^2$ (6s 6p, 5d 6p) & $r_{22}$ & -6833.7 &  \\
$R^1$ (6s 6p, 6p 5d) & $r_{22}$ & -7463.2 &  \\
\end{tabular}
\end{ruledtabular}
\end{table}

\end{document}